\documentclass[preprint,showpacs,preprintnumbers,amsmath,amssymb,superscriptaddress]{revtex4}

\usepackage{graphicx,amsfonts}% Include figure files
\usepackage{epsfig}
\usepackage{dcolumn}% Align table columns on decimal point
\usepackage{bm}% bold math
\usepackage{subfig}

\begin{document}

%\preprint{IFT-P.xx/2009}
%\preprint{ArXiv:yymm.nnnn}
\title{A model with two inert scalar doublets}

% \altaffiliation[Also at ]{Physics Department, XYZ University.}%

\author{A. C. B. Machado}%
\email{ana@ift.unesp.br}
\affiliation{
Instituto  de F\'\i sica Te\'orica--Universidade Estadual Paulista \\
R. Dr. Bento Teobaldo Ferraz 271, Barra Funda\\ S\~ao Paulo - SP, 01140-070,
Brazil
}

\author{V. Pleitez}%
\email{vicente@ift.unesp.br}
\affiliation{
Instituto  de F\'\i sica Te\'orica--Universidade Estadual Paulista \\
R. Dr. Bento Teobaldo Ferraz 271, Barra Funda\\ S\~ao Paulo - SP, 01140-070,
Brazil
}

\date{01/15/16}% It is always \today, today,

% \altaffiliation[Also at ]{Physics Department, XYZ University.}%

\begin{abstract}
We consider an extension of the standard model (SM) with three $SU(2)$ scalar doublets  and a discrete $S_3\otimes \mathbb{Z}_2$ symmetries. The irreducible representation of $S_3$ has a singlet and a doublet, and here we show that the singlet corresponds to the SM-like Higgs and the two additional $SU(2)$ doublets forming a $S_3$ doublet are inert. 
In general, in a three scalar doublet model,  with or without $S_3$ symmetry,  the diagonalization of the mass matrices implies arbitrary unitary matrices. However, we show that in  our model these matrices are of the tri-bimaximal type. We also analyzed the scalar mass spectra and the conditions for the scalar potential is bounded from below at the tree level. We also discuss some phenomenological consequences of the model.
\end{abstract}


\maketitle

\section{Introduction}
\label{sec:intro}

In 2012 it  was discovery at the LHC a neutral spin-0 resonance with properties (mass and couplings) that are compatible, within the experimental error, with those of the scalar SM-Higgs boson~ \cite{atlas:2012gk, cms:2012gu}.
However, on the one hand, there is no experimental evidence confirming that only one of such sort of scalars does exist. An in the other hand, there are experimental evidence, e.g. the existence of  Dark Matter (DM) and neutrinos masses and mixing, that strongly suggest that the SM is not the ultimate theory of nature. In this context, we may need to add new scalars, to play the role of the DM candidate or in order to justify the difference between the mass scale of the neutrinos and the charged leptons. The question is, if there are more scalar doublets, how many of them?  The simplest case is to add one more doublet. This is well motivated because it allows spontaneous $C\!P$ violation if at the same time flavor changing neutral currents (FCNC) are allowed~\cite{Lee:1973iz,Wu:1994ja}. The latter processes strongly constrain
the masses and the mixing angles in the scalar sectors. For a recent review of the phenomenology of the two Higgs doublet models (2HDM) see Ref.~\cite{Branco:2011iw}. The next simple situation is having three doublets in which it is possible to have spontaneous and hard $C\!P$ violation~\cite{Weinberg:1976hu} and at the same time to avoid FCNC if some extra symmetries are introduced. Next, we can introduce more 
Higgs scalar doublets, for instance, it may be motivated by the implementation of the Peccei-Quinn symmetry and the unification of the three interactions, see Ref.~\cite{Dias:2004hy} and references therein.  

Among all these possibilities the case of three doublets with the same quantum number is interesting if we assume that the replica of three generations occurs not only in the fermion sector but also in the scalar sector. However, a general three doublet model (3HDM) has a very complicated scalar potential with six  parameters with dimension of mass ($\mu^2$s), and many dimensionless ones ($\lambda$s). Notwithstanding in physics, when the degrees of freedom augment it motivates the introduction of new symmetries. In fact, to reduce the number of parameters in the scalar sector usually symmetries, like $\mathbb{Z}_2$~\cite{Grzadkowski:2010au}, are introduced.  In some cases one of the scalar doublets is inert. However, the $Z_2$ symmetry still allows four $\mu^2$s and 23 dimensionless real parameters. The possibility of an $S_3$ symmetry is also explored in~\cite{Kubo:2004ps,Das:2014fea}. In the case of the 3HDM with $S_3$ symmetry has only two parameters with dimension of mass and eight dimensionless ones  i.e., 
in terms of the number of parameters the scalar potential in the 3HDM plus a $S_3$ symmetry has almost the same as the general 2HDM. The problem with this model is that in general several possibilities are allowed, and some of them are not physical because they imply the existence of massless physical neutral scalar. Another difficulty  is the existence of FCNC effects~\cite{Chen:2004rr}. Usually also the neutral scalar with mass of 125 GeV is obtained only in the decoupling limit~\cite{Das:2014fea}.  All these effects arise mainly because, i) the mass matrices mix all the scalars in each charge sector and, because of this, the unitary matrices that diagonalize the  respective mass matrices are general ones in each case; ii) an arbitrary vacuum alignment is assumed with all the vacuum expectation values (VEVS) being different.

Here we will consider a 3HDM with an $S_3$ symmetry in which the SM-like scalar is automatically identified without requiring a decoupling limit. This is a consequence of a particular vacuum alignment and the absence of FCNC at tree level is a consequence of this vacuum alignment plus the condition of fermions transforming trivially under the symmetry $S_3$. Another important consequence of this vacuum alignment is that the scalars in the $S_3$ doublet are inert~\cite{Deshpande:1977rw,Barbieri:2006dq}. It means that they do not contribute to the spontaneous symmetry breaking and do not couple to fermions. They interact only with the vector and the other scalar bosons. 

The SM extensions with one inert doublet model (IDM)  as a candidate to dark matter have been already considered in Refs.~\cite{Cirelli:2005uq,LopezHonorez:2006gr,Hambye:2007vf,Cao:2007rm, Andreas:2008xy, Lundstrom:2008ai,Hambye:2009pw,LopezHonorez:2010tb,Krawczyk:2013jta}. However having two inert doublets allows to have a multi-component dark matter scenario~\cite{Daikoku:2011mq, Biswas:2013nn, Bhattacharya:2013hva}, because not only the real scalar and pseudoscalar may be DM as in the IDM, but now we have two real scalar fields and two pseudoscalar ones, each of them may contribute to the DM density. 

The outline of this paper is as follows. In Sec.~\ref{sec:scalar} we give the most general scalar potential involving three
Higgs doublets which is invariant under the gauge and $S_3$ symmetries. We also consider in that section the mass spectra in the
scalar sector when all VEVs satisfy the alignment $v_1=v_2=v_3=v=v_{SM}/\sqrt{3}$ and the singlet ($S$) and the doublet ($D$) of $S_3$ are originated from a triplet i.e., in the reducible representation. In this situation there are mass degenerate states in each scalar sector because a residual $S_2$ symmetry remains. However the mass degeneracy may be lifted by introducing terms that break the $S_2$ symmetry softly.  We dubbed this case A. In Sec.~\ref{sec:modelb} we consider the case when $S = H_1$ and $D = (H_2, H_3)$ with $v_{1} = v_{SM}$ and $v_2 = v_3 = 0$. We call this case B. 
We show that both cases, \textit{before} the spontaneous symmetry breaking (SSB), are related by a weak basis transformation. However, \textit{after}  the SSB both cases are still equivalent but only in the vacuum alignment is considered here. We also consider in this case the situation when the $S_2$ symmetry is softly broken avoiding the mass degeneracy, in this case the equivalence between both cases is also lost. The Yukawa interactions are the same in both cases and are briefly discussed in Sec.~\ref{sec:yukawa}. In Sec.~\ref{sec:potential} we study the positivity of the scalar potential at the tree level, while in Sec.~\ref{sec:pheno} we consider some 
phenomenological consequences. Our conclusions are in Sec.~\ref{sec:con} and in the appendices we show the 
constraint equations for arbitrary VEVs, for case A in Appendix~\ref{sec:a1}, and for case B in Appendix.~\ref{sec:a2}.

\section{Three Higgs-scalar doublet model and $S_3$ symmetry}
\label{sec:scalar}

We present an extension of the electroweak standard model with three Higgs scalars, all of them transforming as doublets under 
$SU(2)$ and having $Y=+1$. Some of them transform under $S_3$ as a doublet $D=(D_1,D_2)\equiv \textbf{2}$,  and some as a singlet $S\equiv\textbf{1}$. As we will see, the latter one is identified with the SM-like Higgs and the former ones are inert.

The most general scalar potential invariant under $SU(2) \otimes U(1)_Y \otimes S_3$ symmetry is given by:
\begin{eqnarray}
V(D,S) &=& \mu^2_sS^\dagger S+\mu^2_d [D^\dagger\otimes  D]_1 +\lambda_1
([D^\dagger\otimes  D]_1)^2
+  \lambda_2 [(D^\dagger\otimes D)_{1^\prime}(D^\dagger\otimes
D)_{1^\prime}]
\nonumber \\ &+&\lambda_3[(D^\dagger \otimes D)_{2^\prime}(D^\dagger\otimes D)_{2^\prime}]_1
+\lambda_4(S^\dagger S)^2+
\lambda_5[D^\dagger\otimes D]_1 S^\dagger  S+[\lambda_6 [[S ^\dagger D]_{2^\prime} [S^\dagger   D]_{2^\prime}]_1\nonumber \\ &+& H.c.]+
\lambda_7 S^\dagger [ D \otimes D^\dagger]_1 S +[\lambda_8[(S^\dagger\otimes D)_{2^\prime}(D^\dagger \otimes D)_{2^\prime}]_1+H.c.]
\label{potential1}
\end{eqnarray}
Denoting an arbitrary doublet by $\textbf{2}=(x_1,x_2)$, we have the product rule\textbf{S} as
$\textbf{2}\otimes\textbf{2}=\textbf{1}\oplus\textbf{1}^\prime\oplus
\textbf{2}^\prime$ where $\textbf{1}=x_1y_1+x_2y_2$,
$\textbf{1}^\prime=x_1y_2-x_2y_1$,
$\textbf{2}^\prime=(x_1y_2+x_2y_1,x_1y_1-x_2y_2$), and
$\textbf{1}^\prime\otimes\textbf{1}^\prime=\textbf{1}$~\cite{Ishimori:2010au}. Let us define $S=(s^+\,s^0)^T$, $D_i=(d^+_i\,d^0_i)^T,\;i=1,2$. In terms of the $S$ and $D_i$ fields, the potential in Eq.~(\ref{potential1}) is written as
\begin{eqnarray}
V(S,D_1,D_2) &=& \mu^2_sS^\dagger S+\mu^2_d (D^\dagger_1D_1+D^\dagger_2 D_2) +\lambda_1
(D^\dagger_1D_1+D^\dagger_2 D_2)^2
+  \lambda_2 (D^\dagger_1D_2-D^\dagger_2D_1)^2
\nonumber \\ &+&\lambda_3[(D^\dagger_1D_2+D^\dagger_2D_1)^2+(D^\dagger_1D_1-D^\dagger_2D_2)^2]
+\lambda_4(S^\dagger S)^2+
\lambda_5  (D^\dagger_1D_1+D^\dagger_2 D_2)S^\dagger  S \nonumber \\&+&[\lambda_6(S^\dagger D_1S^\dagger D_1+S^\dagger D_2S^\dagger D_2)+H.c.]
+
\lambda_7 S^\dagger (D_1D^\dagger _1+D_2 D^\dagger_2) S \nonumber \\&+& \lambda_8[S^\dagger D_1(D^\dagger_1D_2+D^\dagger_2 D_1)+S^\dagger D_2(D^\dagger_1D_1-D^\dagger_2D_2)+H.c.]
\label{potential2}
\end{eqnarray}
Notice that the potential is written in terms of the symmetry eigenstates independently of how we form the singlet and the doublet.
If $\mu^2_d>0$ only the singlet $S$ gain a VEV and if $\lambda_8=0$ this vacuum is stable at tree and the one-loop level.
For this term be forbidden we impose a $\mathbb{Z}_2$ symmetry under which $D\to -D$ and $S$ and all the other fields are even. 
However, in the appendix we consider the constraint equations with a general vacuum alignment in order to study under what conditions we have  $\langle D\rangle=0$ and we find that independently of the signal of $\mu^2_d$, it is possible to have the vacuum alignment considered in this paper.
The three-Higgs scalar potential has already been considered in the literature in Refs.~\cite{Kubo:2004ps,Das:2014fea,Mondragon:2007af,Beltran:2009zz}  but not in the inert doublets context. In fact, unlike the present paper, all these articles have used the $S_3$ symmetry to address the texture of the fermion mass matrices using a general vacuum alignment. 

\section{Case A}
\label{sec:modela} 
 
Let us now consider the case when the three scalar doublets are in the  reducible triplet representation of $S_3$, say, $\textbf{3}= (H_1,H_2,H_3)$ where $H_i=(H^+_i\,H^0_i)^T$. This reducible 
representation is broken down to the irreducible singlet and doublet ones, i.e., $\textbf{3}~=~\textbf{2}
+\textbf{1}\equiv D+S$, where:
\begin{eqnarray}
&&S=\frac{1}{\sqrt3}(H_1+H_2+H_3)\sim\textbf{1},\nonumber \\&&
D\equiv (D_1,D_2)=\left[\frac{1}{\sqrt6}(2H_1-H_2-H_3),\frac{1}{\sqrt2}(H_2-H_3)\right]\sim\textbf{2},
\label{ma}
\end{eqnarray}
or, explicitly in terms of the symmetry eigenstates $H^{+,0}_i,\,i=1,2,3$
\begin{eqnarray}
&& S\equiv\left(\begin{array}{c}
s^+\\s^0
\end{array} 
\right) =\frac{1}{\sqrt3}\, \left(\begin{array}{c}
H^+_1+H^+_2+H^+_3\\
H^0_1+H^0_2+H^0_3
\end{array} 
\right),
D_1\equiv\left(\begin{array}{c}
d^+_1\\d^0_1
\end{array} 
\right) =\frac{1}{\sqrt6}\, \left(\begin{array}{c}
2H^+_1-H^+_2-H^+_3\\
2H^0_1-H^0_2-H^0_3
\end{array} 
\right),\nonumber \\&&
D_2\equiv\left(\begin{array}{c}
d^+_2\\d^0_2
\end{array} 
\right) =\frac{1}{\sqrt2}\, \left(\begin{array}{c}
H^+_2-H^+_3\\
H^0_2-H^0_3
\end{array} 
\right).
\label{defn2}
\end{eqnarray}
The decomposition of the symmetry eigenstates we make as usual, as $H^0_i\!~=~\!(1/\sqrt{2})(v_i\!~+~\!\eta^0_i\!~+~\!i\,a^0_i),\;i=1,2,3$. We assume for the sake of simplicity that the VEVs are real, however see Sec.~\ref{sec:pheno}. 
The general constraint equations for the case when all VEVs are different from zero are given in the Appendix~\ref{sec:a1}, Eq.~(\ref{vinculos1}). When $v_1=v_2=v_3=v$ these constraint equations  are reduced to a simple equation:
\begin{equation} 
t_1 =t_2=t_3  =v(\mu^2_s  +    3\lambda_4 v^2),
\label{vinculos}
\end{equation}
and if $t_i=0$ we have $\mu^2_s= - 3 \lambda_4v^2 =- \lambda_4v_{SM}^2 <0$, which implies that $\lambda_4>0$.

All scalar mass square  matrices have the form
\begin{equation}
%\label{massmatrixgeral}
M^2_n=\left(\begin{array}{ccc} a_n &  b_n& b_n  \\ b_n & a_n &  b_n  \\ b_n & b_n  & a_n\end{array}\right),
\label{mss1}
\end{equation}
where $a_n,b_n>0$ and $n$ denotes the scalar sector: $n=h,a,c$ for the scalar, pseudo-scalar and charged scalar 
fields, respectively.

This type of matrix is diagonalized  by an orthogonal matrix, $U_{TBM}$: 
$U^T_{TBM}M^2_nU_{TBM}=\textrm{diag}(a_n + 2b_n,a_n-b_n,a_n-b_n)$, with
$a_n+2b_n\geq0$ and $a_n-b_n\geq0,\;\forall n$, the $U_{TBM}$ is given by
\begin{equation}
U_{TBM} =\left(
\begin{array}{ccc}
\frac{1}{\sqrt3} & - \sqrt{\frac{2}{3}} & 0\\
\frac{1}{\sqrt{3}}& \frac{1}{\sqrt{6}}& - \frac{1}{\sqrt{2}}\\
\frac{1}{\sqrt{3}}& \frac{1}{\sqrt{6}} & \frac{1}{\sqrt{2}}
\end{array}
\right).
\label{mis1}
\end{equation}

In the case of $C\!P$-even neutral scalars, we have
$3 a_h=  2 \mu_d^2 + (2 \lambda_4 +  \bar{\lambda}^\prime) v_{SM}^2$,
and $ 6 b_h= -2 \mu_d^2 + (4 \lambda_4 - \bar{\lambda}^\prime) v_{SM}^2$, where  $\bar{\lambda}^\prime=(\lambda_5
+ \lambda_7 + 2 \lambda_6)$, and the eigenvalues are the following:
\begin{eqnarray}
&&m^2_{h_1}\equiv m^2_h= 2 \lambda_4v^2_{SM}, \nonumber \\&&
m^2_{h_2}=m^2_{h_3}\equiv m^2_H=\mu^2_d+\frac{1}{2}\bar{\lambda}^\prime v^2_{SM}=\mu^2_d+\frac{1}{4}\,\frac{\bar{\lambda}^\prime}{\lambda_4}m^2_h,
\label{mrs1}
\end{eqnarray}
where we have used $v=v_{SM}/\sqrt3$. %Hereafter we will use $h\equiv h_1$ to denote the SM-like Higgs doublet.

Denoting as $h^0_i$ the mass eigenstates, we have $h^0_i=\sum_i(U^T_{TBM})_{ij}\eta^0_j$, where $U_{TBM}$
is given in (\ref{mis1}). Explicitly we have
\begin{equation}
\left( \begin{array}{c}
h^0\\ h^0_2\\ h^0_3
\end{array}
\right)=
\left(\begin{array}{c}
\frac{1}{\sqrt3}(\eta^0_1+\eta^0_2+\eta^0_3)\\
-\frac{1}{\sqrt6}(2\eta^0_1-\eta^0_2-\eta^0_3)\\
-\frac{1}{\sqrt2}(\eta^0_2-\eta^0_3)
\end{array}\right) 
\equiv \textrm{Re} \left(
\begin{array}{c}
s^0\\-d^0_1\\-d^0_2
\end{array}
\right),
\label{obax} 
\end{equation}
and the scalar $h^0\equiv \textrm{Re}\,s^0$ which, in Sec.~\ref{sec:yukawa}, will be identified with the SM Higgs boson and the doublet $h_1\equiv h$ with the SM scalar doublet.

In the $C\!P$-odd neutral scalars sector, the mass matrix is given as in Eq.~(\ref{mss1}) but now with
$ 3 a_a= 2 \mu_d^2 + \bar{\lambda}^{\prime \prime} v_{SM}^2$
and $ 6 b_a= -2 \mu_d^2 - \bar{\lambda}^{\prime \prime} v_{SM}^2$, where  $\bar{\lambda}^{\prime \prime}=(\lambda_5 + 
\lambda_7 - 2 \lambda_6)$ and
in this case we obtain the following masses:
\begin{eqnarray}
&& m^2_{A_1}=0, \nonumber \\&&
m^2_{A_2}=m^2_{A_3}\equiv m^2_A=\mu^2_d+\frac{1}{2} \bar{\lambda}^{\prime \prime} v^2_{SM}
=\mu^2_d+\frac{1}{4}\,\frac{\bar{\lambda}^{\prime\prime}}{\lambda_4}m^2_h .
\label{mps1}
\end{eqnarray}
Denoting $A^0_i$ the pseudo-scalar mass eigenstates, we have $A^0_i=\sum_i(U^T_{TBM})_{ij}a^0_j$ and making the same as in 
Eq.~(\ref{obax}) we obtain that $A^0= \textrm{Im}\,s^0$ is the would-be Goldstone boson, while $A^0_2=-\textrm{Im}\,d^0_1$ and $A^0_3=-\textrm{Im}\,d^0_2$
are physical $C\!P$ odd fields.

Similarly in the charged scalars sector we use Eq.~(\ref{mss1}) with
$6a_c= 2 \mu_d^2 + \lambda_5 v_{SM}^2$ and $12b_c= -2 \mu_d^2  -  \lambda_5 v_{SM}^2$  and in this case we obtain
the following masses:
\begin{eqnarray}
&&m^2_{c_1}=0,\nonumber \\&&
m^2_{c_2}=m^2_{c_3}\equiv 2m^2_c=\mu^2_d+\frac{\lambda_5}{2}v^2_{SM}=\mu^2_d+\frac{1}{8}\,\frac{\lambda_5}{\lambda_4}m^2_h,
\label{mcs1}
\end{eqnarray}
and, if $H^+_i$ denote the charged scalar symmetry eigenstates and $h^+_i$ the respective mass eigenstates,
we have $h^+_i=\sum_i(U^T_{TBM})_{ij}H^+_j$. Using again the Eq.(\ref{mis1}) we obtain $s^+=h^+ $ the would-be the charged Goldstone boson, and the physical charged scalars: $-d^+_1=h^+_2$ and $-d^+_2=h^+_3$. 

We summarize these results by using mixing matrix in Eq.~(\ref{mis1}), and writing the Higgs scalars 
doublet $D$ and the singlet $S$, but now in terms of the mass eigenstates, $h^0_i,A^0_i$ and $h^\pm_i$, as
\begin{eqnarray}
S\equiv \phi=\left(\begin{array}{c}
h^+\\ \frac{1}{\sqrt2}(\sqrt{3}v+h^0+iA^0)\end{array}\right),\; D\equiv -(\phi_1,\phi_2),\; \phi_k=\left(\begin{array}{c}
h^+_k\\ \frac{1}{\sqrt2}(h^0_k+iA^0_k)\end{array}\right),
\label{a11}
\end{eqnarray}
where $k=2,3$. 

We have the sum rule from Eqs.~(\ref{mrs1}),~(\ref{mps1}),~(\ref{mcs1}):
\begin{equation}
m^2_H+m^2_A+2m^2_c=3\mu^2_d+\frac{1}{\lambda_4}(\bar{\lambda}^\prime+\bar{\lambda}^{\prime\prime}+\lambda_5)m^2_h.
\label{sumrule}
\end{equation}

The mass degeneracy in Eqs. (\ref{mrs1}), (\ref{mps1}) and (\ref{mcs1}), is due to a residual symmetry as we will see below. The $\mu^2_d$ parameter appearing in these equations is not related to the spontaneous 
symmetry breaking. Thus, since $\mu^2_d$ is not protected by any symmetry, it may be larger than the
electroweak scale. On one hand, if $\mu^2_d> v^2_{SM}$ (assuming the $\lambda's$ are of order one) the masses of  the  scalar
$h^0_{2,3}$, pseudo-scalar $A^0_{2,3}$ and the charged scalar $h^\pm_{2,3}$ are heavier than $h^0$, independently
of the values of the $\lambda$'s and $v_{SM}$. On the other hand, if $\mu^2_d<0$ and $\bar{\lambda}^\prime, 
\bar{\lambda}^{\prime \prime}>0,\lambda_5>0$, all these particles may be lighter than $h^0$. However, since $\lambda_5$ may be negative, $\lambda_4$ is always positive, and in this case $m^2_c<0$, as can be seen from Eq.~(\ref{mcs1}), here we will consider only $\mu^2_d>0$ and larger than $\vert(\lambda_5/8\lambda_4)\vert m^2_h$.

The potential in Eq.~(\ref{potential2}) can be written in terms of $SU(2)$ scalar doublets 
with their components being the mass eigenstates given in Eq.~(\ref{a11}):
\begin{eqnarray}
V(\phi_i)&=&3\lambda_4 v^2\phi^\dagger \phi+\mu^2_d(\phi^\dagger_1\phi_1+\phi^\dagger_2\phi_2)+\lambda_1(\phi^\dagger_1\phi_1+\phi^\dagger_2\phi_2)^2
+\lambda_2(\phi^\dagger_1\phi_2-\phi^\dagger_2\phi_1)^2\nonumber \\&+&\lambda_3[(\phi^\dagger_1\phi_2+\phi^\dagger_2\phi_1)^2+(\phi^\dagger_1\phi_1-
\phi^\dagger_2\phi_2)^2]+\lambda_4(\phi^\dagger \phi)^2+\lambda_5\phi^\dagger \phi (\phi^\dagger_1\phi_1+\phi^\dagger_2\phi_2)\nonumber\\
&+&\lambda_7 [ \vert \phi^\dagger \phi_1\vert^2 + \vert \phi^\dagger \phi_2\vert^2] + \{\lambda_6 [(\phi^\dagger \phi_1)^2+(\phi^\dagger_2\phi)^2]+
\lambda_8[\phi^\dagger  \phi_1(\phi^\dagger_1 \phi_2+\phi^\dagger_2 \phi_1)\nonumber\\&+&\phi^\dagger \phi_2(\phi^\dagger_2 \phi_2-\phi^\dagger_1 \phi_1)]+H.c.\}.
\label{potential3}
\end{eqnarray}
Notice that this scalar potential is the same as that in Eq.~(\ref{potential2}). However the later one was written in terms of the symmetry eigenstates and (\ref{potential3}) is in terms on the mass eigenstates.
This  occurs only in this model and not in any 3HDM and it is a consequence of the $S_3$ symmetry and the vacuum alignment.

Notice that this scalar potential with three Higgs scalar doublets under $SU(2)$, is as simple as the two 
doublet case, see for instance in Ref.~\cite{Gunion:2002zf}. From Eq.~(\ref{potential3}), we can see that if $\lambda_8=0$ there is still a residual 
$S_2$ symmetry: it is invariant under the exchange of the doublets $\phi_1\leftrightarrow \phi_2$. Notice, however, that the mass 
degeneracy is due to the fact that the $\lambda_8$ term does not contribute to the Higgs scalar masses, this is easy to be verified, once $\lambda_8$ corresponds only to the trilinear and quartic interactions among the three doublets. Anyway, we have considered only the $\lambda_8=0$ case due to the $\mathbb{Z}_2$ symmetry considered above.

We will show later on under which conditions the potential  in Eq.~(\ref{potential2}) (or (\ref{potential3})) is bounded from below. For the 
moment, just notice that when $v_1=v_2=v_3$,  if $\lambda_4>0$, the minimum of the scalar potential 
($V_{min}=-\lambda_4v^4_{SM}$) is global and stable minimum if  the masses square, given in (\ref{mrs1}), (\ref{mps1}), and (\ref{mcs1}) are all positive and, if the conditions for the $\lambda$'s given in Sec.~\ref{sec:potential} are satisfied. However, the stability of the solution $v_1=v_2=v_3$ under radiative corrections will be studied elsewhere.

The residual $S_2$ symmetry can be broken, if necessary, to avoid the mass degeneracy and also
the domain wall problem. This can be done by quantum corrections~\cite{Goudelis:2013uca} and/or by soft terms in the 
scalar potential. As an illustration, here we break this symmetry by adding the following quadratic terms 
$\mu^2_{nm}H^\dagger_nH_m$, $n,m=2,3$ to the scalar potential in (\ref{potential1}). The mass matrices in all the scalar 
sectors are now of the form
\begin{equation}
M^2_n=\left(\begin{array}{ccc} a_n & b_n& b_n  \\ b_n & a_n+\mu^2_{22} &  b_n+\mu^2_{23}  \\ b_n & b_n+
\mu^2_{23}  & a_n+\mu^2_{33}\end{array}\right),
\label{mss2}
\end{equation}
where $\mu^2_{nm}$  are naturally small, and we will assume that are real for the sake of simplicity. Although when 
$ \mu^2_{22} =  \mu^2_{33}=  \nu^2$ and $\mu^2_{23}  = \mu^2$ the matrix above is still diagonalized by the tribimaximal 
matrix, as the neutrinos masses~\cite{Altarelli:2010fk}, this is not possible with scalars fields: in this case there is no would-be
Goldstone bosons. In order to have the correct number of these bosons we have to impose that 
$\mu^2_{22}=\mu^2_{33}=-\mu^2_{23}\equiv \mu^2$.  In this case the matrix in Eq.~(\ref{mss2}) is still diagonalized by 
tribimaximal matrix in Eq.~(\ref{mis1}), and the eigenvalues are now $(2a_n + b_n,a_n - b_n, a_n - b_n+\mu^2)$ and we 
still have $S=\phi$ and $D=-(\phi_1,\phi_2)$, as in the previous case.

\section{A change of weak basis: Case B}
\label{sec:modelb}

We can build the singlet and a doublet of $S_3$ with just one $SU(2)$-doublet, say $H_1$, 
and the other  two, say $H_2$ and $H_3$, transform as the irreducible doublet of $S_3$, i.e.,
\begin{equation}
S = H_1\sim\textbf{1} , \quad D = (H_2, H_3) \sim\textbf{2}.
\label{mb}
\end{equation}
But, note that, the two bases are related by the tribimaximal matrix in Eq.~(\ref{mis1}), i.e, 
\begin{equation}
\left(\begin{array}{c}
S\\D_1\\D_2
\end{array} 
\right)=U^T_{TBM}\left(\begin{array}{c}
H_1\\H_2\\H_3
\end{array} 
\right)
\label{basis}
\end{equation}
with $U^T_{TBM}$ being the transpose of the matrix in Eq.~(\ref{mis1}) and $S$, $D_1$ and $D_2$ here are those in Eq.~(\ref{ma}). The representation in Eq.~(\ref{mb}), is called here case B. It was considered since a long time ago \cite{Pakvasa:1977in, Mondragon:2007af,Bhattacharyya:2010hp,Teshima:2012cg,Kaneko:2006wi} but in other context and different motivations. 

Although both cases in (\ref{ma}) and (\ref{mb})  are related by the transformation in Eq.~(\ref{basis}) and can be considered just the same model in two different basis, we can see that this is true only \textit{before} the spontaneous symmetry breaking. The VEV of the $S_3$ triplet in case A is given by
\begin{equation}
\langle \left(
\begin{array}{c}
H_1\\H_2\\H_3
\end{array} 
\right)\rangle
=\left(
\begin{array}{c}
v_1\\v_2\\v_3
\end{array} 
\right).
\label{ufa}
\end{equation} 
When the decomposition in Eq.~(\ref{ma}) is used and the vacuum alignment $v_1=v_2=v_3=\frac{v_{SM}}{\sqrt3}$ is used it
implies the inert character of the doublet $D$. However, in case B we have
\begin{equation}
\langle \left(
\begin{array}{c}
H_1\\H_2\\H_3
\end{array} 
\right)\rangle
=\left(
\begin{array}{c}
v_{SM}\\0\\0
\end{array} 
\right).
\label{ufa2}
\end{equation} 
We see that the vacua in (\ref{ufa}) and (\ref{ufa2}) are related by the transformation in (\ref{basis}) only when $v_1=v_2=v_3=\frac{v_{SM}}{\sqrt3}$.
Hence, only in this situation both cases are identical \textit{before} and \textit{after} the spontaneous symmetry breaking. 
But in a general vacuum the inert character of the doublet is lost because in this case the mass matrices are of the general form, a full $3\times3$ matrix, and after the field rotation the SM-Higgs like will be a linear combination of this three fields, this implies that at tree level and / or loop level all scalars couple to all fermions.Hence we have to impose in case B that $v_2=v_3=0$. In fact, the constraint equations are different and are given in the Appendix \ref{sec:a2}. These constraint equations are the same, see Eq.~(\ref{vinculos}) with $3v^2=v^2_{SM}$ only in the vacuum alignment used in this paper.

The constraint equations in Eq.~(\ref{vinculos2}) implies, with the vacuum alignment given above, $\mu^2_s= -\lambda_4v^2_{SM}$ and the mass square matrices are all diagonal: there is no mixing among the scalar fields in each charge sector. At tree level the masses are the same as in case A, see Eqs.~(\ref{mrs1}) - (\ref{mcs1}). The doublets of $SU(2)$ written in terms of the mass eigenstates are denoted, as before, by $\phi$ and $\phi_{1,2}$. In this case the scalar potential in terms of these fields is given also in Eq.~(\ref{potential3}) again with $\lambda_8 = 0$, this shows that even after the SSB the models are equivalent. Unlike the case A, there is no mixing among the mass eigenstate scalar fields therefore these fields are in  the irreducible representations of $S_3$ too: $S \equiv \phi$ and $D=(d_1,d_2)\equiv (\phi_1,\phi_2)$. 

The transformation $\phi_1\leftrightarrow \phi_2$ is again a residual $S_2$ symmetry which, if necessary, can be softly broken by adding terms like $\mu^2H^\dagger_2H_3$,  ($\mu^2$ is also considered to be real for 
simplicity). In this case, the mass matrices are given by
\begin{equation}
M^2_n=\left(
\begin{array}{ccc}
m^2_{n_1} &0& 0\\
0&m^2_{n_2}&\mu^2\\
0&\mu^2& m_{n_2}
\end{array}
\right),
\label{massmb}
\end{equation}
where $n=h,a,c$, for scalar, pseudo-scalar and charged scalar field, respectively. The mixing now is only in the inert
sector and the masses square are
\begin{eqnarray}
&&\bar{m}^2_{h_1}=m^2_h,\;\; \bar{m}^2_{h_2}=m^2_h-\mu^2,\;\;
\bar{m}^2_{h_3}=m^2_h+ \mu^2,\nonumber \\&&
\bar{m}^2_{a_1}=0,\;\; \bar{m}^2_{a2}=m^2_a-\mu^2, \;\;\bar{m}^2_{a3}=
m^2_a+\mu^2,\nonumber \\&&
\bar{m}^2_{c_1}=0,\;\;\bar{m}^2_{c_2}=m^2_c-\mu^2,\;\; \bar{m}^2_{c_3}=m^2_c+\mu^2,
\label{massfb}
\end{eqnarray}
where $m^2_h,m^2_a$ and $m^2_c$ are given in Eqs.~(\ref{mrs1}), (\ref{mps1}), and (\ref{mcs1}), respectively. 
The mass matrices of the form in (\ref{massmb}) are diagonalized by the orthogonal matrix
\begin{equation}
U = \left(
\begin{array}{ccc}
1 &                                               0 & 0 \\
0 &-\frac{1}{\sqrt{2}}  & \frac{1}{\sqrt{2}} \\
0 &  \frac{1}{\sqrt{2}} &  \frac{1}{\sqrt{2}}
\end{array}
\right),
\label{uva}
\end{equation}
and the mixing between $\phi_1$ and $\phi_2$ is maximal.

Thus, in terms of the mass eigenstate fields, the scalar doublets of $SU(2)$ are written as $S=\phi$ and
$D\equiv -(D_1,D_2)=-(-\phi_1+\phi_2,\phi_1+\phi_2)$, where $\phi_i$ are the $SU(2)$ doublets written in terms of the
mass eigenstate fields. Explicitly
\begin{eqnarray}
&&S\equiv \phi =\left(\begin{array}{c}
h^+\\ \frac{1}{\sqrt2}(v_{SM}+h^0+iA^0)\end{array}\right),\; \nonumber \\&&
\phi_1= \left(\begin{array}{c}
\frac{1}{\sqrt2}(-h^+_2+h^+_3)\\ \frac{1}{2}[-h^0_2+h^0_3+i(-A^0_2+A^0_3)]\end{array}\right),
\phi_2= \left(\begin{array}{c}
\frac{1}{\sqrt2}(h^+_2+h^+_3)\\\frac{1}{2}[ h^0_2+h^0_3+i(A^0_2+A^0_3)]\end{array}\right).
\label{a13}
\end{eqnarray}
It is important to note again that after the degeneracy breaking we lose the connection between the two cases as can be seen by comparing Eq.~(\ref{a11}) with (\ref{a13}).

\section{The Yukawa sector}
\label{sec:yukawa}
If in cases A and B in the lepton and quark sectors all fields transform as singlet under $S_3$, they only 
interact with the singlet $S$ as following:
\begin{equation}
-\mathcal{L}_{yukawa}=\bar{L}_{iL}(G^l_{ij}l_{jR}S+G^\nu_{ij}\nu_{jR}\tilde{S})+ \bar{Q}_{iL}(G^u_{ij}u_{jR}
\tilde{S}+G^d_{ij}d_{jR}S) +H.c.,
\label{yukawaint}
\end{equation}
$\tilde{S}=i\tau_2S^*$ and we have included right-handed neutrinos.

We see that the fermion masses, as in the SM, arise only through the  VEV of the singlet $S$ which is the only field, 
or linear combination of fields, with a non-zero VEV, see Eq.~(\ref{a11}) and (\ref{a13}). Hence, there is no FCNC in the lepton 
and quark sectors at the tree level. Moreover, we obtain arbitrary mass matrices from Eq.~(\ref{yukawaint}), because 
there is just one source of the fermion masses which are given by $M^f=(v_{SM}/\sqrt2)G^f$, $f=l,\nu,u,d$ and where 
$v_{SM}=246$ GeV. The neutral interactions are $(\sqrt2/v_{SM})\bar{f}_L\hat{M}^ff_R h^0$, where $\hat{M}^f$ is the 
diagonal mass matrix in the $f$-sector. These mass matrices are general enough to accommodate a realistic 
$V_{PMNS}$ and $V_{CKM}$ mixing matrices. Moreover, since the right-handed neutrinos may have a Majorana mass term 
we can have a type-I seesaw mechanism. 

Notwithstanding, unlike the case of the SM, having only one source of fermion masses is not guaranteed to avoid FCNC in the scalar sector. 
In fact, the case of natural flavor conservation when there are discrete symmetries was not considered in Ref.~\cite{Glashow:1976nt}. Hence, it is worth to considering briefly this issue. Let $\mathcal{D}$ be a 
generic non-Abelian discrete symmetry with multiplication law $*$ under which the left- and right-handed fermions, namely $f_{\mathcal{D}_L}$ and $f_{\mathcal{D}_R}$,  
are in different representations of the gauge symmetry  but are  
singlet under the $\mathcal{D}$ symmetry. The scalar multiplets, $H_{\mathcal{D}}$, transform non-trivially under the gauge symmetry, but are singlet of $\mathcal{D}$ since this is the scalar that couple to fermions. The Yukawa 
interactions are of the form $\bar{f}_{\mathcal{D}_L}*f_{\mathcal{D}_R}*H_{\mathcal{D}}\sim\textbf{1}$, i.e.,
it is invariant under the gauge and discrete transformations. Even if 
$H_{\mathcal{D}}$ is in the trivial representation of $\mathcal{D}$ as we have assumed, without the vacuum 
alignment discussed above there are FCNC in each charge sector. With an arbitrary vacuum alignment the relation in Eq.~(\ref{obax}), which implies that $\textrm{Re}S^0=h^0$,  is no longer valid and $S^0$ is a linear component of the three neutral mass eigenstates and all of them contribute to the fermion mass matrices.
It suggests that the vacuum alignment can be added to the 
conditions in Ref.~\cite{Glashow:1976nt} to have natural flavor conservation in neutral currents at the tree level when discrete symmetries are present in the model. Here we have considered $\mathcal{D}=S_3$. 

\section{Analysis of the scalar potential}
\label{sec:potential}

The scalar potential has to be bounded from below to ensure its stability. In the SM this is easy at least at tree level, we just have to ensure that the quartic term in the potential has $\lambda > 0$. In theories that increase the number of scalar multiplets it is more difficult to ensure the positivity of the potential in all directions.  A scalar potential has a quadratic form in the quartic couplings in the form $A_{ab} \xi^2_a \xi^2_b $.  If the matrix $A_{ab}$ is copositive in the sense of Ref.~\cite{Kannike:2012pe}, it is possible to ensure that the potential is bounded from below. Let us apply this analysis to our case.

We obtain the copositive conditions in the quartic terms in the scalar potential given in Eq.~(\ref{potential3}), by defining:
\begin{equation}
\label{def1}
|\zeta_{i}|^2 =  \xi_{i}^2,\quad 
\zeta^\dagger_{i} \zeta_{j}   =  \xi_{i} \xi_{j} \rho_{i} e^{i \theta_{i}}
\end{equation}
where $\zeta_i=S,D_1,D_2$, and  $\rho_{i}$ and $\theta_{i}$ are not physical parameters. From the scalar potential of Eq.~(\ref{potential2}) we obtain the  matrix $A$ in the base $(\xi^2_1, \xi_{2}^2, \xi_{3}^2)$ the matrix elements are given by:
\begin{eqnarray}
A_{11} & = & \lambda_4, \nonumber \\
A_{22} & = & \lambda_1 + \lambda_3, \nonumber \\
A_{33} & = & A_{22}, \nonumber \\
A_{12} & = & A_{21} =  \frac{1}{2} [\lambda_5 + \rho^2_{1} (\lambda_7 + 2 \lambda_6 \cos(2 \phi_1))], \nonumber \\
A_{13} & = & A_{31} = \frac{1}{2} [\lambda_5 + \rho^2_{2} (\lambda_7 + 2 \lambda_6 \cos(2 \phi_2))], \nonumber \\
A_{23} & = & A_{32} =  2(\lambda_3 - \lambda_2) +\rho_3^2 (\lambda_2 + \lambda_3) \cos(2 \phi_3).
\end{eqnarray}
Now we most minimize the potential with respect to the free parameters $\rho_{i}$ and $\phi_{i}$. For the terms $2 \lambda_7 \cos(2 \phi_1)$ and $2 \lambda_7 \cos(2 \phi_2)$ it is obvious that the minimum will be when  $\cos(2 \phi_1)= \cos(2 \phi_2) = -1$, for the element $A_{23}$ to the minimum occurs for $\rho_3 = 1$ and  $\cos(2\phi_3) = -1$, which leaves us with

\begin{eqnarray}
A_{11} & = & \lambda_4, \nonumber \\
A_{22} & = & \lambda_1 + \lambda_3, \nonumber \\
A_{33} & = & A_{22} \nonumber \\
A_{12} & = & A_{21} =  \frac{1}{2} (\lambda_5 + \rho^2_{1} (\lambda_7 - 2 \lambda_6 ,) \nonumber \\
A_{13} & = & A_{31} = \frac{1}{2} (\lambda_5 + \rho^2_{2} (\lambda_7 - 2 \lambda_6 ), \nonumber \\
A_{23} & = & A_{32} = - 4 \lambda_2.
\end{eqnarray}
Now if $\lambda_7 - 2 \lambda_6 \geq  0$, the minimum of the potential is obtained by setting $\rho_1 = \rho_2 =0$, but if $\lambda_7 - 2 \lambda_6 \leq  0$ then the minimum is given by $\rho_1 = \rho_2 =1$. To simplify our analysis  and since the results for $\rho_1$ and $\rho_2$ are equal we will set $\rho_1 = 0$ and $\rho_2 = 1$. Finally we have the following expressions for the matrix elements,

\begin{equation}
A=\left(\begin{array}{ccc}
\lambda_4 & \frac{1}{2} \lambda_5 & \frac{1}{2} [\lambda_5 + \lambda_7 - 2 \lambda_6 ]\\
& \lambda_1 + \lambda_3 &  - 4 \lambda_2\\
& & \lambda_1 + \lambda_3
\end{array} 
\right). 
\label{matrixa}
\end{equation}

For a symmetric matrix A of order 3 the copositivity criteria are summarized as follows: $a_{ii} > 0$ and $v_{ij}= a_{ij} + \sqrt{a_{ii} a_{jj}} > 0$ and $\sqrt{a_{11}a_{22} a_{33}} + a_{12}\sqrt{a_{33}} +  a_{13}\sqrt{a_{22}} +  a_{23}\sqrt{a_{11}} + \sqrt{v_{12} v_{13} v_{23} } > 0$.
Explicitly we obtain:
\begin{eqnarray}
&&\lambda_4  >  0,
\nonumber  \\  &&
\lambda_1 + \lambda_3  >  0 ,
\nonumber  \\  &&
\lambda_5   + 2 \sqrt{\lambda_4(\lambda_1 + \lambda_3)}  >  0, 
\nonumber \\  &&
\lambda_5 + \lambda_7  - 2 \lambda_6 + 2 \sqrt{\lambda_4(\lambda_1 + \lambda_3)}  >  0 , 
\nonumber  \\ && 
\lambda_1 + \lambda_3  >  4 \lambda_2, 
\label{nossa1}
\end{eqnarray} 
and
\begin{eqnarray} 
&&
(\lambda_1 - 2 \lambda_2 - \lambda_3) \sqrt{\lambda_4} + \sqrt{(- 4 \lambda_2) (\lambda_1 + \lambda_3) \lambda_4} 
\nonumber \\  &+& 
  \sqrt{\lambda_1 + \lambda_3} (\lambda_5 + \lambda_7 - 2 \lambda_6) +
\frac{  \sqrt{\lambda_1 - \lambda_2}  (2 ~\sqrt{(\lambda_1 + \lambda_3) \lambda_4} + \lambda_5 + \lambda_7 - 2 \lambda_6)}{\sqrt{2}} > 0,
\nonumber \\  &&
(- 4 \lambda_2) \sqrt{\lambda_4} + \sqrt{(- 4 \lambda_2) (\lambda_1 + \lambda_3) \lambda_4} 
\nonumber \\  &+& 
  \sqrt{\lambda_1 + \lambda_3} (\lambda_5 ) + 
\frac{  \sqrt{\lambda_1 - \lambda_2} ( 2~ \sqrt{(\lambda_1 + \lambda_3) \lambda_4} + \lambda_5)}{\sqrt{2}} > 0.
\label{nossa2}
\end{eqnarray}

It is easy to verify that if the conditions in Eqs.~(\ref{nossa1}) are satisfied the conditions in Eq.~(\ref{nossa2}) are automatically satisfied. Hence, the positivity of the scalar potential is guarantee just by the conditions in Eq.~(\ref{nossa1}).

\section{Some phenomenological consequences}
\label{sec:pheno}

It is well known that two-Higgs doublet models have an interesting phenomenology~\cite{Branco:2011iw}. For instance,
i) in a broad class of this type of models  there is $C\!P$ violation arising purely from the 
exchange of Higgs bosons but FCNC are allowed~\cite{Lee:1973iz}, and ii) in the class of models with inert scalars 
the lightest  neutral fields is, at least in some range of the parameters, a dark matter candidate in the universe. Here we will consider only these two phenomenological aspects in this model. The first one is $C\!P$ violation and, secondly the possibility of having a dark matter candidate.

In general in three Higgs doublet models there is also $C\!P$ violation via de exchange of scalar fields~\cite{Weinberg:1976hu}. We will analyze this issue in the present model.
In case A, and the potential in Eq.~(\ref{potential1}) or in Eq.~(\ref{potential2}). Let us suppose that the VEVs are complex, and still imposing $v_1e^{i\theta_1}=v_2e^{i\theta_2}=v_3e^{i\theta_3}=
Ve^{i\Theta}$ as a stable minimum of the scalar potential. The phase $\Theta$, which appears only in the singlet $S$,  
can be transform away with a global $U(1)$ transformation as it happens  in the standard model. On the other hand, 
if $\theta_1\not=\theta_2\not=\theta_3$ we lost the inert feature of the two $SU(2)$ doublets in $D=(\phi_1,\phi_2)$. Thus, if we want two inert doublets there is no spontaneous $C\!P$ violation through the VEVs. We can also consider the possibility to have hard explicit $C\!P$ violation 
through complex coupling  constants in the scalar potential because $\lambda_6$ may be complex, we can define $\lambda_6 = \vert \lambda_6 \vert e^{i \alpha_6}$. In this case, it is possible to transformed away the $\lambda_6$ phase by making the global phase rotations $S \rightarrow S e^{i a_S}$  and $D \rightarrow D e^{i a_D}$, and choosing $a_D - a_S = \alpha_6/2$, the $\lambda_6$ phase can be eliminated, wherefore we see that in this context there is no $C\!P$ violation in the scalar sector, just the hard violation in the quark and lepton mixing matrices.

We can try, also, to have soft explicit $C\!P$ violation through 
the quadratic  non-diagonal term in the scalar potential $\mu^2h^\dagger_2h_3$ assuming that $\mu^2$ is complex, as in Ref.~\cite{Wu:1994ja}. However it is not possible in case A once the mass matrices in Eq.~(\ref{mss2}) are not diagonalized by tribimaximal-type matrix and the inert feature of the two extra doublets is lost. However, this source of $C\!P$ violation is possible in case B since, as can be seen from Eq.~(\ref{massmb}), notwithstanding
the mixing and the $C\!P$ violation occurs only in the inert sector. 

It is well known that there exists a range of the parameters in which an inert doublet is a candidate for dark matter (DM)~\cite{Barbieri:2006dq, Goudelis:2013uca,Cirelli:2005uq,LopezHonorez:2006gr}.
This may also imply, in the present model,  
contributions to the invisible decay of the SM-like Higgs~\cite{Belanger:2013kya,Mambrini:2011ik,Chatrchyan:2014tja}.
It has been shown in Ref.~\cite{Krawczyk:2013jta} that, in the context of one 
inert Higgs doublet (IDM), there are three allowed regions of masses that are compatible with observed value of $\Omega_{DM}h^2$:
i) $\stackrel{<}{\sim} 10$ GeV; ii) 40-150 GeV, and iii)  $\stackrel{>}{\sim} 500$ GeV. Notice that the regions i) 
and that in 40-60 GeV there is SM Higgs invisible decay. 

The same may happen in the present model with $h^0_{2,3}$. Here we will only show that, for a range of the parameters
and for the three allowed regions above the spin-independent cross section for the $h^0_{2,3}$-nucleon sca\-tte\-ring agrees with the Xenon100 
results~\cite{Aprile:2012nq}, the Lux results~\cite{Akerib} and the theoretical prediction of Xenon1T. And at the same time for the region i) and 40-60 GeV, $h^0\to h^0_{2,3}$ may be compatible with the invisible width decay. Here we will consider only when there is mass degeneracy in case A.

The spin-independent cross section for DM-nucleon scattering is given by~\cite{Krawczyk:2013jta}:
\begin{equation}
\sigma_{SI} = 2\times \frac{m_p^4 \;\bar{\lambda}^{\prime \, 2}  f^2}{4 \pi (m_p + m_{h_2^0})^2 m^4_{h_2^0}},
\label{si}
\end{equation}
where the factor 2 is because we have two mass degenerated inert scalars, and $f=0.326$, see~\cite{Mambrini:2011ik}; 
and the invisible Higgs width by:
\begin{equation}
\Gamma (h^0 \rightarrow h^0_2h^0_2(h^0_3h^0_3)) =2\times  \frac{\bar{\lambda}^{\prime \, 2} v_{SM}^2}{32 \pi m_{h_2^0}}
\sqrt{1 - \left(\frac{4m_{h_2^0}}{m_{h}}\right)^2}.
\label{inv}
\end{equation}

\begin{figure}[tbp]
\centering % \begin{center}/\end{center} takes some additional vertical space
\subfloat[]{\includegraphics[width=9cm,height=6cm]{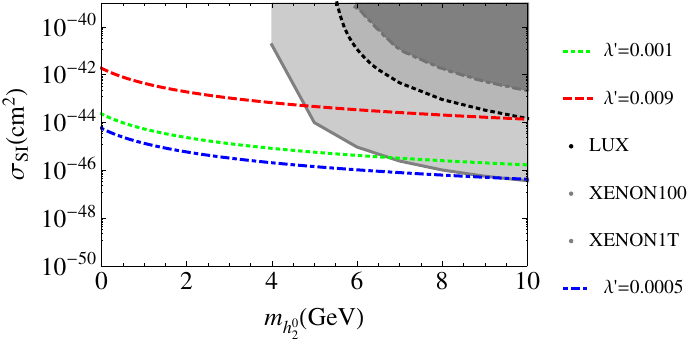}}\\
\subfloat[]{\includegraphics[width=9cm,height=6cm]{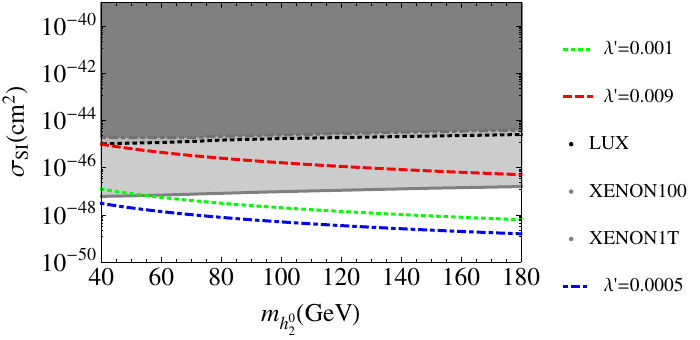}} \\
\subfloat[]{\includegraphics[width=9cm,height=6cm]{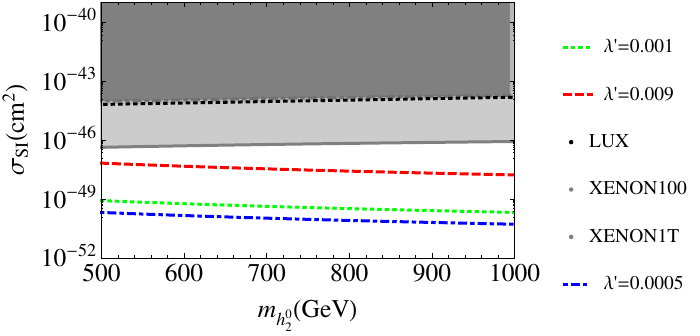}} \\
\caption{\label{fig1} The gray areas show the regions excluded by Xenon, Lux and the theoretical prediction Xenon1T for $\sigma_{SI}$ defined in Eq.~(\ref{si}) as a function of the DM-candidate mass, and for three different values of $\bar{\lambda}^\prime$. (a) is the region for masses less than 10 GeV, (b) shows the behavior of Eq.~(\ref{si}) for masses between 40 and 160 GeV, finally in (c) we show the allowed region for masses larger than 500 GeV.}
\end{figure}

In Fig.~\ref{fig1} we show the excluded region given by Xenon100~\cite{Aprile:2012nq} and Lux \cite{Akerib} results and the theoretical prediction for Xenon1T, Fig.~\ref{fig1}(a) shows the behavior of Eq.~(\ref{si})  as a function of the masses for a fixed $\bar{\lambda}^\prime$ for masses less than 10 GeV, in this case the best solution is for $\bar{\lambda}^\prime = 5 \times 10^{-4}$, but with masses lower than 6 GeV all values are allowed. Fig.~\ref{fig1}(b) shows the behavior of Eq.~(\ref{si})  as a function of $\bar{\lambda}^\prime$ for masses between 40 and 160 GeV, in this case we have two good solution for $\bar{\lambda}^\prime = 5 \times 10^{-4}$ for the entire range and $\bar{\lambda}^\prime = 10^{-3}$ for masses between 60 and 160 GeV. Finally in Fig.~\ref{fig1}(c) we show that for masses larger than 500 GeV $\bar{\lambda}^\prime$  is allowed for a range between $5 \times 10^{-4}$ and $9 \times 10^{-3}$. These values are in agreement with the calculation of the relic density for this model as shown in Ref.~\cite{Fortes:2014dca}.

\begin{figure}[tbp]
\centering % \begin{center}/\end{center} takes some additional vertical space
\subfloat[]{\includegraphics[width=.75\textwidth,origin=c,angle=0]{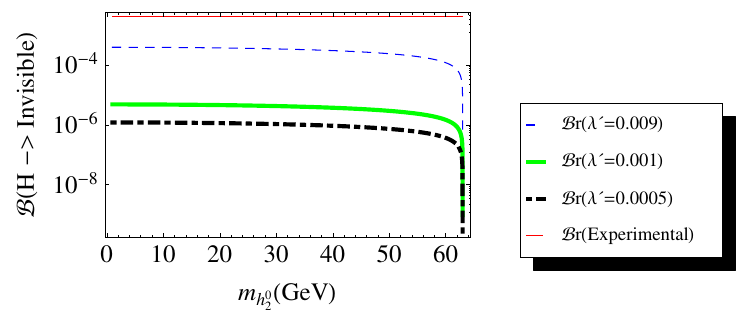}}\\
\subfloat[]{\includegraphics[width=.75\textwidth,origin=c,angle=0]{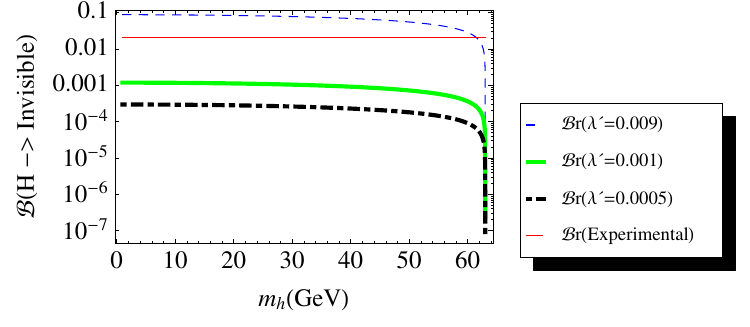}}
\caption{\label{fig2} The invisible Higgs width (a) defined in Eq.~(\ref{inv}), and the branching ratio (b), $Br(h~\to~\textrm{inv})~=~\frac{\Gamma(h \to \textrm{inv})}{\Gamma_{SM}+\Gamma(h \to \textrm{inv})}$ as functions of the scalar mass with $\bar{\lambda}^\prime = 5 \times 10^{-4}$, $\bar{\lambda}^\prime = 9 \times 10^{-3}$ and $\bar{\lambda}^\prime = 10^{-3}$ for masses in the range $m_h <$ 62 GeV. As can be seen $\bar{\lambda}^\prime= 9 \times 10^{-3}$ is excluded by data, if we want that the scalar be a dark matter candidate.}
\end{figure}

In Fig.~\ref{fig2}(a) we show the invisible Higgs width, using Eq.~(\ref{inv}), and in Fig.~\ref{fig2}(b) the branching ratio 
$Br(h~\to~\textrm{inv})~=~\frac{\Gamma(h \to \textrm{inv})}{\Gamma_{SM}+\Gamma(h \to \textrm{inv})}$, as functions 
of the scalar mass and with three values for $ \bar{\lambda}^\prime =  5 \times 10^{-4}, 1 \times 10^{-3}$ and $9 \times 10^{-3}$ for the mass range $m_{h_2^0}~ <~ 62$ GeV. Note that the curve for 
$\bar{\lambda}^\prime=0.009$ is excluded if we want to have a dark matter candidate and 
the invisible branching ratio $\mathcal{B}(h \rightarrow DM) < 0.2$~\cite{Belanger:2013kya}. 

Since $\bar{\lambda}^\prime=\lambda_5+\lambda_6+2\lambda_7$, 
from the constraints in Eq.~(\ref{nossa1}) and from the expressions for the masses in Eq.~(\ref{mrs1}) and Eq.~(\ref{mcs1}),  we have 
\begin{equation} 
\frac{2 (m_H^2 - \mu_d^2)}{v_{SM}^2 \sqrt{\lambda_4 (\lambda_1 + \lambda_3)}}~> - 1~,~\frac{ (4m_c^2 - 2 \mu_d^2)}{v_{SM}^2 \sqrt{\lambda_4 (\lambda_1 + \lambda_3)}}~> - 1.
\label{extra}
\end{equation} 

\begin{figure}[!ht]
\begin{center}
\includegraphics[width=7cm,height=5cm]{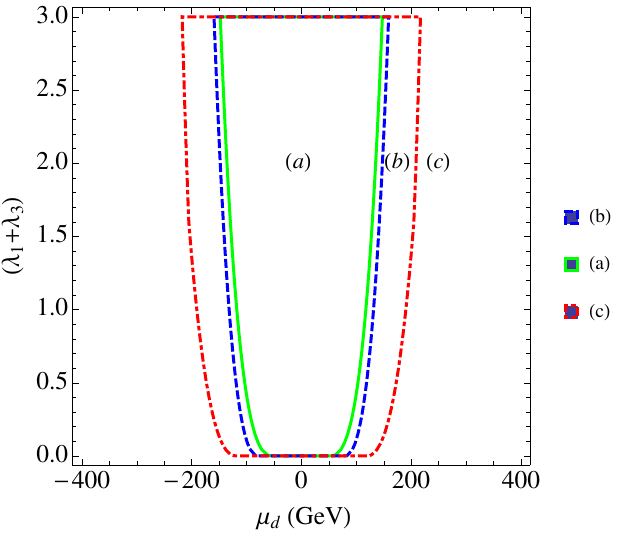}\\
\caption{\label{fig3} The allowed region, using Eq.~(\ref{extra}), for $\mu_d^2$ and $(\lambda_1 + \lambda_3) $ when we fix (a) $m_{H}^2 = 54.1$ and $m_{c}^2 = 85$, (b)$m_{H}^2 = 80$ and $m_{c}^2 = 95$, and (c) $m_{H}^2 = 168$ and $m_{c}^2 = 84.7$. With $\lambda_4=0.13$ fixed by the SM Higgs mass.}
\end{center}
\end{figure}

From  Eq.~(\ref{extra}), we obtain the allowed region for $\mu_d^2$ and $(\lambda_1 + \lambda_3) $ if we fix $m_{H}^2$ and $m_{c}^2$, and $\lambda_4=0.13$ is fixed by the SM Higgs mass. These are shown in Fig.~\ref{fig3} for a) $m_{H}^2 = 54.1$ and $m_{c}^2 = 85$, b) $m_{H}^2 = 80$ and $m_{c}^2 = 95$, and c) $m_{H}^2 = 168$ and $m_{c}^2 = 84.7$. These values are also compatible with the experimental data of dark matter, as was shown in Ref.~\cite{Fortes:2014dca}.  

\begin{figure}[!ht]
\begin{center}
\subfloat[]{\includegraphics[width=7cm,height=5cm]{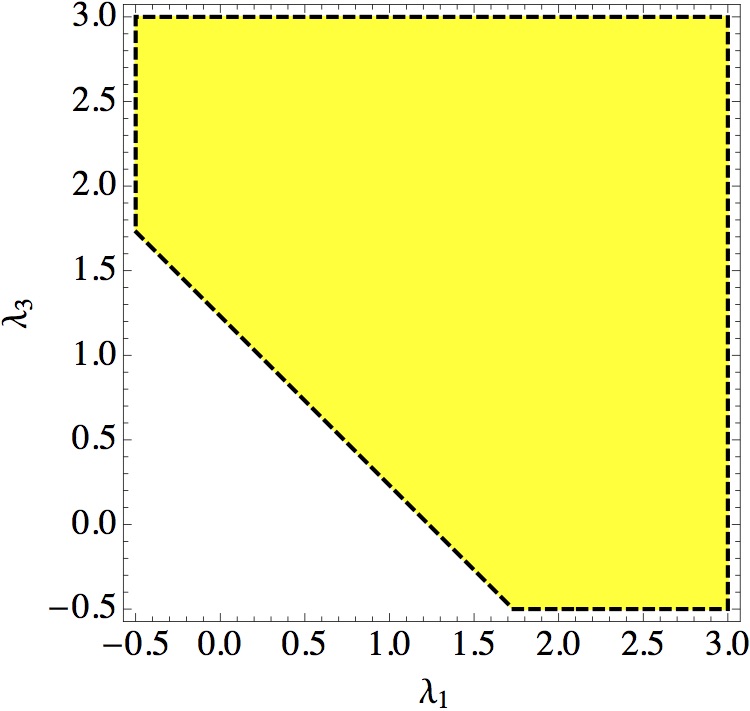}}\caption{\label{fig4} 
Using Eq.~(\ref{nossa1}), $\lambda_5   >- \sqrt{\lambda_4(\lambda_1 + \lambda_3)}$ , and the results in Fig.~\ref{fig3} we obtain the allowed region for $\lambda_1$ and $ \lambda_3 $ when we fix $\lambda_5 = -0.4$ and $\lambda_4=0.13$ is fixed by the SM Higgs mass, is the yellow area in figure above.}
\end{center}
\end{figure}

The presence of two inert doublet implies in contributions for  $h^0\to\gamma\gamma$~\cite{Cardenas:2012bg}, and $h^0\to Z\gamma$~\cite{Fortes:2014dia}. In the latter paper it was obtained the best value for $\lambda_5$, that fit the current data for $h \rightarrow \gamma \gamma$, when it is $\lambda_5 = -0.4$, in Fig.~\ref{fig4} we show the constraints on $\lambda_1$ and $\lambda_3$ using the third line of Eq.~(\ref{nossa1}), ($\lambda_5 + \sqrt{\lambda_4(\lambda_1 + \lambda_3)} > 0$).

\section{Conclusions}
\label{sec:con}

If the Higgs sector has, as in the fermion sector, three sequential generations, we should expect the existence of
extra symmetries to make the interactions and the mass spectra simplest in the scalar sector. This is because, 
in general, three Higgs doublet models have  complicated scalar potentials and mass matrices in each charge sector are diagonalized by arbitrary unitary $3\times3$ matrices having each one three mixing angles and six phases (some of them may be absorbed). In the present model because of the $S_3$ symmetry and the vacuum alignment, the entries of the rotation matrices  are, at the tree level, Glebsch-Gordan-like coefficients. This eliminate plenty of new parameters that should have  to be determined by experiments.  In fact, the scalar potential in this models is as simple as that in a general two Higgs doublet model. The only difference is the $\lambda_8$ term in the scalar potential, see Eq.~(\ref{potential1}) or (\ref{potential2}). Anyway it is necessary that $\lambda_8=0$ in order to maintain the inert character of the doublet of $S_3$, $D$.

Moreover, like multi-Higgs models with no flavor changing neutral currents mediated by
neutral scalars, the only mixing parameters appearing in the fermion charged interactions are the CKM and PMNS 
angles and phases. For more details see  Ref.~\cite{Cardenas:2012bg}. We would like to stress that the existence of two inert doublets, and the flavor conservation in the neutral currents mediated  by scalars are consequences of three ingredients: i) the $S_3$ symmetry, ii)  the representation content of the fermion and scalar multiplets under $S_3$, and, iii) the vacuum alignment. 

If the lightest neutral scalars are the $CP$-even as we have assumed here, the $CP$-odd ones can be produced at LHC in vector-boson fusion. 
This also deserves a detailed study.

\acknowledgments

The authors would like to thank to thank  CAPES under the PNPD program (ACBM) for fully support and to CNPq for the partial support  (VP).
for partial support.

\appendix

\section{Constraint equations in model A}
\label{sec:a1}

Expanding the scalar potential in Eq. (\ref{potential1}) as a function of VEV's, we will obtain:

\begin{eqnarray}
V & = & \frac{1}{36} (6 \mu_s^2 (v_1 + v_2 + v_3)^2  + 
   12 \mu_d^2 (v_1^2 + v_2^2 - v_2 v_3 + v_3^2 - v_1 (v_2 + v_3)) 
   \\ \nonumber &+&
   4 \lambda_1 (v_1^2 + v_2^2 - v_2 v_3 + v_3^2 - v_1 (v_2 + v_3))^2 + 
   4 \lambda_3 (v_1^2 + v_2^2 - v_2 v_3 + v_3^2 - v_1 (v_2 + v_3))^2) \\ \nonumber &+&
    \lambda_4 (v_1 + v_2 + v_3)^4 + 2 \lambda_5 (v_1 + v_2 + v_3)^2 (v_1^2 + v_2^2 - v_2 v_3 + v_3^2 - 
      v_1 (v_2 + v_3)) 
      \\ \nonumber &+& 
   2 \lambda_6 (v_1 + v_2 + v_3)^2 (v_1^2 + v_2^2 - v_2 v_3 + v_3^2 - 
      v_1 (v_2 + v_3)) 
      \\ \nonumber &+&
   4 \lambda_7 (v_1 + v_2 + v_3)^2 (v_1^2 + v_2^2 - v_2 v_3 + v_3^2 - 
      v_1 (v_2 + v_3)) \\ \nonumber &-&
  2 \sqrt{2}   \lambda_8 (v_1 + v_2 - 2 v_3) (2 v_1 - v_2 - v_3) (v_1 - 2 v_2 + v_3) (v_1 + v_2 + 
      v_3),
 \end{eqnarray}
the constraint equations are explicitly given by:
\begin{eqnarray}
18t_1&=&6\mu^2_d(2v_1\!\!-\!\!v_2\!\!-\!\!v_3)+6\mu^2_sV+2(\Lambda_1\!\!-\!\!4\sqrt{2}\lambda_8)v^3_1
-[(\Lambda_2\!\!+\!\!\sqrt{2}\lambda_8)(3v^2_1+v^2_2+v^2_3)
\nonumber \\&&
-(\Lambda_3\!\!-\!\!7\sqrt{2}\lambda_8)v_2v_3](v_2\!\!+\!\!v_3)
\!+\!6[(\Lambda_4\!\!+\!\!2\sqrt{2}\lambda_8)(v^2_2\!\!+\!\!v^2_3)
\!+\!(\Lambda_5\!\!-\!\!2\sqrt{2}\lambda_8)v_2v_3]v_1
\label{vinculos1}
\end{eqnarray}

\begin{eqnarray}
18t_2&=&-\mu^2_d(v_1-2v_2+v_3)+6\mu^2_sV+ 2\Lambda_1v^3_2+(\Lambda_2-\sqrt{2}\lambda_8)(v^3_1+v^3_3+3v^2_2v_3+3v^2_2v_1)\nonumber \\&&
+6(\Lambda_4+2\sqrt{2}\lambda_8)(v^2_1+v^2_3)v_2
+3(\Lambda_5-2\sqrt{2}\lambda_8) (v_1v_3+2v_2v_3+v^2_3)v_1
\label{vinculos2}
\end{eqnarray}

\begin{eqnarray}
18t_3&=&-6\mu^2_d(v_1+v_2-2v_3)+\mu^2_sV+2\Lambda_1 v^3_3
-(\Lambda_2+\sqrt{2}\lambda_8)(v^3_1+v^3_2+3v_1v^2_3+3v_2v^2_3)\nonumber \\&&
+6(\Lambda_4+2\sqrt{2}\lambda_8)(v^2_1+v^2_2)v_3+3(\Lambda_5-2\sqrt{2}\lambda_8)(v_1v_2+v^2_2+2v_2v_3)v_1\nonumber \\&&
\label{vinculos3}
\end{eqnarray}

where
\begin{eqnarray}
&&\Lambda_1=2\lambda^\prime+\lambda_4+2\bar{\lambda}^\prime,\quad
\Lambda_2=2\lambda^\prime -2\lambda_4-\bar{\lambda}^\prime,\quad \Lambda_3=2(\lambda^\prime+2\lambda_4-2\bar{\lambda}^\prime),
\nonumber \\&&
\Lambda_4=\lambda^\prime+\lambda_4,\quad
\Lambda_5=2\lambda_4-\bar{\lambda}^\prime.
\label{defnovas}
\end{eqnarray}

Although the $\lambda_8$ have allows solutions, in this case the $\lambda_8$ symmetry has to be forbiden because it induces a tadpole that destabilize the vacuum alingment.

\section{Constraint equations in model B}
\label{sec:a2}

Expanding the scalar potential in Eq. (\ref{potential2}) as a function of VEV's we have
\begin{eqnarray}
V & = & 
\frac{1}{4} (2 \mu_s^2 v_1^2 + \lambda_4 v_1^4 + 
   v_2^2 (2 \mu_d^2 + (\lambda_5 + \lambda_6 + 2 \lambda_7) v_1^2   
   \\ \nonumber &-&
      2 \lambda_8 v_1 v_2 + (\lambda_1 + \lambda_3) v_2^2) + (2 \mu_d^2 + (\lambda_5 + \lambda_6 + 
         2 \lambda_7) v_1^2   
         \\ \nonumber &+&
          6 \lambda_8 v_1 v_2 + 2 (\lambda_1 + \lambda_3) v_2^2) v_3^2 + (\lambda_1 + 
      \lambda_3) v_3^4)
 \end{eqnarray}

With the representation in Eq.~(\ref{mb}), the constrain equation are
\begin{eqnarray}
2t_1&=&v_1\left[2\mu^2_s+2\lambda_4v^2_1+\bar{\lambda}^\prime (v^2_2+v^2_3)-\frac{\lambda_8}{v_1}\left(v^3_2
+v_2v^2_3\right)\right],\nonumber \\2t_2&=&v_2\left[ 2\mu^2_d+\bar{\lambda}^\prime v^2_1+(\lambda_1+\lambda_3)
(v^2_2+v^2_3)-3\lambda_8\left(v_1v_2-\frac{v_1v^2_3}{v_2}\right)\right],
\nonumber \\2t_3&=&v_3[2\mu^2_d+\bar{\lambda}^\prime v^2_1+2(\lambda_1+\lambda_3)(v^2_2+v^2_3)+6\lambda_8v_1v_2],
\label{vinculos2}
\end{eqnarray}
and we see that even in the general case when $v_1\not=v_2\not=v_3$ they are different from the respective equations in model A, see Eq.~(\ref{vinculos1}).

Notice that the $\lambda_8$ term avoid the zero solution for $v_1$ and $v_2$. If this term is forbidden with a $Z_2$ symmetry under which $D\to-D$ and all the other fields being even under this symmetry, we can have the
solution $v_1=v_{SM}$ and $v_2=v_3=0$.

%%%%%%%%%%%%%%%%%%%%%%%%%%%%%%%%%%%%%%%%%%%%%%%%%%%%%%%%%%%%%%%%

\newpage

\end{document}